# How is Vaping Framed on Online Knowledge Dissemination Platforms?

Keyu Chen[1], Yiwen Shi[2], Jun Luo[3], Joyce Jiang[3], Shweta Yadav[4],
Munmun De Choudhury[7], Ashique R. Khudabukhsh[6], Marzieh Babaeianjelodar[1],
Frederick L. Altice[1], and Navin Kumar[1,8]

[1]Yale University School of Medicine, New Haven, CT 06510, USA
[2]Yale University, New Haven, CT 06510, USA
[3]University of California, Los Angeles
[4]University of Illinois Chicago
[5]The University of Edinburgh, College of Arts
[6]Rochester Institute of Technology
[7]Georgia Institute of Technology
[8]National University of Singapore

**Abstract.** Studying how vaping is framed on various knowledge dissemination platforms (e.g., Quora, Reddit, Wikipedia) is central to understanding the process of knowledge dissemination around vaping. Such understanding can help us craft tools specific to each platform, to dispel vaping misperceptions and reinforce evidence-based information. We analyze 1,888 articles and 1,119,453 vaping posts to study how vaping is framed across multiple knowledge dissemination platforms (Wikipedia, Quora, Medium, Reddit, Stack Exchange, wikiHow). We use various NLP techniques to understand these differences. For example, cloze tests and question answering results indicate that Quora is an appropriate venue for those looking to transition from smoking to vaping. Other platforms (Reddit, wikiHow) are more for vaping hobbyists and may not sufficiently dissuade youth vaping. Conversely, Wikipedia may exaggerate vaping harms, dissuading smokers from transitioning. A strength of our work is how the different techniques we have applied validate each other. Based on our results, we provide several recommendations. Stakeholders may utilize our findings to design informational tools to reinforce or mitigate vaping (mis)perceptions online.

## 1 Introduction

The recent introduction of alternative forms of nicotine products into the marketplace (e.g., e-cigarettes/vapes, heated tobacco products, and smokeless tobacco) has led to a more complex informational environment [11]. The scientific consensus is that vape aerosol contains fewer numbers and lower levels of toxicants than smoke from combustible tobacco cigarettes [18]. Among youth in the USA, adolescent nicotine vaping use increased from 2017 to 2019 but then started declining in 2020 [14]. Among adults, a Cochrane review found that nicotine vapes probably do help people to stop smoking for at least six months, working better than nicotine replacement therapy and nicotine-free e-cigarettes [5]. Given that vaping is represented in the public health environment both as a smoking cessation tool and harm to youth health, it is highly controversial and polarizing, with inconsistent messaging across various platforms [7]. For example, manufacturers, retailers, and social media influencers have claimed that e-cigarettes contain only water vapor and are harmless [1]. Such messaging may downplay the risks of vape use and be in part responsible for the youth vaping epidemic [4]. Conversely, there also exists messaging that vapes are just as or more harmful than smoking [13], perhaps deterring current cigarette smokers who are unable to quit from transitioning to vaping. Similarly, regarding the outbreak of vaping-related lung injury (EVALI), most cases were related to consumption of vitamin E acetate, an additive included in some tetrahydrocannabinol devices. However, news reports have not always differentiate between tetrahydrocannabinol devices and standard nicotine-based vapes, perhaps disproportionately characterizing vaping harms. Such vaping-related news may have triggered national and state-level policy responses, and influenced public perceptions (including misperceptions) regarding the harms of vaping [8].

While there has been research around vaping perceptions on news and social media, there is limited analysis of how vaping is framed on digital knowledge dissemination platforms [15]. By knowledge dissemination platforms, we refer to platforms such as social question-and-answer sites (e.g., Quora), social news aggregation sites (e.g., Reddit), online encyclopedias (e.g., Wikipedia), and online publishing platforms (e.g., Medium). Such platforms are where individuals obtain health information, discuss products (e.g., vapes) and personal health, or get quick answers to health questions. Studying how vaping is framed on these platforms is central to understanding the process of knowledge dissemination around vaping. Such understanding can help us craft tools specific to each platform, to dispel vaping misperceptions and reinforce evidence-based information. For example, findings can lead to tools that buttress accurate vaping information on Wikipedia with peer-reviewed literature, but correct misperceptions on Quora. Such targeted tools may aid in reducing youth vaping and improving smoking cessation rates.

Despite the significance of the problem noted above, existing research studying knowledge dissemination platforms' framing of vaping is limited. Most research around vaping on online platforms centers on responses to the 2020 outbreak of vaping-related lung injury (EVALI) [10, 12], or content analysis of vaping. In this paper, we demonstrate how vaping is framed on multiple knowledge dissemination platforms. Our main research question (RQ) is as follows: How is vaping framed in various knowledge dissemination platforms? Our findings suggest that some platforms (Medium, Quora, Stack Exchange) are appropriate for individuals seeking tobacco harm reduction information. Other platforms (Reddit, wikiHow) are more for vaping hobbyists and may not sufficiently dissuade youth vaping. Conversely, Wikipedia may exaggerate vaping harms, dissuading smokers from transitioning. Stakeholders may utilize our findings to design informational tools to reinforce or mitigate vaping (mis)perceptions online.

## 2    Data and Method

**Data** We first selected three content experts who had published at least ten peer-reviewed articles in the last three years around vaping. The context experts separately developed lists of knowledge dissemination platforms most relevant to vaping. Each expert developed a list of ten platforms independently, and we selected only platforms common to all three experts' lists: Wikipedia, Quora, Medium, Reddit, Stack Exchange, wikiHow. Examples of platforms not selected: Facebook, LinkedIn, ChaCha, Answers.com. To capture vaping-related text on these platforms, we used queries based on a related systematic review [2]: electroniccigarette, electronic cigarette, electronic cig, e-cig, ecig, e cig, e-cigarette, ecigarette, e cigarette, e cigar, e-juice, ejuice, ejuice, e-liquid, eliquid, e liquid, e-smoke, esmoke, e smoke, vape, vaper, vaping, vape-juice, vape-liquid, vapor, vaporizer, boxmod, cloud chaser, cloudchaser, smoke assist, ehookah, e-hookah, e hookah, smoke pod, e-tank, electronic nicotine delivery system.
We collected 50 articles on Wikipedia, with the Wikipedia API. For Quora, we used the Pushshift API and obtained 5209 questions and 33493 answer posts. Regarding Reddit, we used the Pushshift API to obtain 132258 posts from 2006 to 2022. For Stack Exchange, we collected 413 posts. We collected 17 wikiHow pages. For Medium we collected 1820 articles. We then removed posts around vaping cannabis as this was not relevant to our research question. We removed posts and articles where the title/question contained the following: *weed, marijuana, cannabis, pot, joint, blunt, mary jane*, resulting in the final dataset: Quora (4890 questions and 31890 answer posts); Reddit (129092 posts and 953168 comments); Medium (1820 articles). Two reviewers independently examined 10% of the remaining articles or posts within each dataset to confirm salience with our research question. The reviewers then discussed their findings and highlighted items deemed relevant across both lists, determining that 85% were relevant.

**Question Answering** Question answering can help us understand how different platforms *answer* the same questions about vaping, perhaps revealing differences in vaping frames. For example, Wikipedia may be more likely to present vaping harms compared to Reddit. We used BERT [3] for answer extraction. The model was applied separately on data from each platform, except wikiHow, which had insufficient data. Questions were developed based on input from content experts. Each content expert first developed a list of ten questions separately. The three experts then discussed their lists to result in a final list of four questions that were broadly similar across all three original lists, and final questions are as follows: What is vaping for? What are the advantages of vaping compared to smoking cigarettes? Why are teens vaping? What is the biggest concern with vaping? We highlighted one question at a time and fed it to the model. While we would have preferred to use more than four questions for our question answering analysis, only four questions were agreed upon by the content experts. This is largely due to disagreement among content experts as to what questions should be included, largely resulting from the controversial nature of vaping, and that academics are in disagreement about the harms and merits of vaping [6]. The model extracts answers for the question leveraging on context information in each article or post. To stay within the admitted input size of the model, we clipped the length of the text (title + body text) to 512 tokens. Each question provided one answer per article or post. We randomly sampled 500, 1000, 1500, and 2000 answers per question. We found that a random sample of 1000 answers provided the greatest range and quality of answers, assessed by two reviewers (80% agreement). We thus randomly sampled 1000 answers per question and content experts then selected the top 10, where possible, most representative answers per question for each platform.

**Cloze Tests** We used BERT and cloze tests to understand the differences between vaping framing across platforms. Cloze tests represent a fill-in-theblank task given a sentence with a missing word. For example, *winter* is a likely completion for the missing word in the following cloze task: In the [MASK], it snows a lot. We developed several cloze tests with input from content experts. Each content expert first developed a list of ten cloze tests separately. Then the three experts compared their lists to only retain items appearing in both lists, resulting in a final list of four cloze tests: i) The main issue with vaping is [MASK]; ii) The worst thing about vaping is [MASK]; iii) Teens like vaping because it's [MASK]; iv) Vaping is [MASK] to our health. We applied BERT on each platform's dataset, where possible, to identify the differences in the top five results for each cloze test for Reddit and Quora. There was insufficient data on other platforms to perform similar analyses.

**Translation across Platforms through Large-scale Language Models** Next, we use large-scale language models to understand the differences between posts across platforms. Such models serve a range of purposes. We use these models to perform single word translation where the model takes a word in a source language as input and outputs an equivalent word in a target language [9]. For example, in a translation system performing English to Spanish translation, if the input word is hello, the output word will be ola. We build on earlier work [9] and treat platforms as different languages. As our *languages* are actually English from different platforms, on most occasions, translations will be identical. As an example, *food* in English used by the Reddit users (Reddit-English) will likely translate into the same in Quora-English [9]. The interesting cases are pairs where translations do not match. The output is not inherently misaligned, and the algorithm simply produces word pairs. We determine whether there is a misalignment through human review. Most of the time, pairs will match (aligned). However, sometimes the pairs will not match (misaligned) and this is of interest. An example pair that may not match in our context is *ingredients,* *additives*. *Ingredients* may be used in favorable contexts in Reddit-English, much like how *additives* may be used in unfavorable contexts within Quora-English. Thus, while both words have different meanings and representations in each platform, they are treated the same by the translation algorithm, creating a mismatch in translation for *ingredients* and *additives*. Such word pairs are misaligned pairs. Such mismatches can provide insights on the differences in how vaping is framed between Reddit and Quora. We fed the models our posts, divided by platform (Reddit, Quora), as two different languages. There was insufficient data on other platforms to conduct similar analyses.

We provide a brief technical overview of the technique used, drawing from [9]. Let D1 and D2 be two monolingual text corpora authored in languages L1 and L2 respectively. With respect to D1 and D2, V1 and V2 denote the source and target vocabularies. A word translation scheme that translates L1 to L2 takes a source word (W1) as input and produces a single word translation W2 (more details in [9]). A translation algorithm [16] drives this process. The algorithm requires two monolingual corpora and a bilingual seed lexicon of word translation pairs as inputs. First, two separate monolingual word embeddings are induced using a monolingual word embedding learning model. FastText was used to train monolingual embedding. Next, a bilingual seed lexicon is used to learn an orthogonal transformation matrix, which is then used to align the two vector spaces. Finally, to translate a word from the source language to the target language, we multiply the embedding of the source word with the transformation matrix to align it with the target vector space. Then, the nearest neighbour of the aligned word vector in the target vector space is selected as the translation of the source word in the target language. Two reviewers manually inspected the top 5000 salient translation pairs, ranked by frequency [9], between Reddit and Quora. Reviewers were instructed to independently order the list with most mismatched pairs at the top. By most mismatched we refer to pairs with the greatest difference in meaning, such as *ingredients, additives*. Examples of less mismatched pairs are those which are different words but closer in meaning, such as *got, started*, and *liquids, juices*. The reviewers then compared the top 15 most mismatched pairs in their lists to look for items common to both lists. Two pairs were common to both lists, and are displayed in the results section. Examples of pairs not selected are *combustible, carcinogenic*, *addicted, pointless*, and *industry, government*. As a clarification, our goal in using techniques described in [9] was not to provide an improvement over an existing technique, but to demonstrate the technique in a different context. While we largely used the work of [9] unchanged, we calculated similarity scores between sentences to find illustrative examples of misaligned pairs between Quora and Reddit where the pairs appear in highly similar contexts - essentially sentences that have similar meanings but with different words. Similarity scores were calculated with Sentence-bert, a modification of the pretrained BERT network that uses siamese and triplet network structures to derive semantically meaningful sentence embeddings that can be compared using cosine-similarity.

## 3    Results

**Overview** We first provide an overview of our data. Examples of popular Reddit posts (most upvotes, comments) were *New Vape Trick! What is going on with vaping and lung disease? Instead of banning vaping, maybe parent your own kids instead of asking the government to do your job for you*. Posts generally centered on content for vapers and concerns around vaping regulation. Posts catered to vapers, with limited anti-vaping content. Examples of popular Quora questions (most comments, upvotes, shares) include *Are e-cigarettes or vapes a safer alternative to tobacco products? Vaping is bad for you and I'm trying to get my friend Emma who is 13 to stop How do I do this? If you smoke would you consider vaping instead or are there particular reasons why you won't? Have you tried it and gone back to smoking or does it satisfy all your cravings?*

Quora comments are increasing over time, with three comments in 2012, and 7,799 in 2021. Quora seems to focus on vaping as a possible tool for smoking cessation, and deterring youth from vaping. Overall, it seems to be the most balanced platform around promoting vaping as a smoking cessation tool and limiting youth use. Medium articles with the most engagement (*claps*, responses) were *Vaping information/myths debunked, How Juul Exploited Teens' Brains to Hook Them on Nicotine, If You're Still Vaping, Experts Urge You to Stop, Strict E-Cigarette Laws Could Send Smokers Back to the Real Thing, What's Known and Not Known About the Mysterious Vaping Illness*. Medium tends to have a broad scope of vaping viewpoints, with anti-vaping articles, and articles targeted at vaping enthusiasts. The number of articles and amount of engagement increased over time till 2019 and then rapidly declined, likely due to attention around EVALI. Examples of popular Stack Exchange posts (most upvotes or views) are *Are there rules for vaping etiquette? What is the cause of the vapingrelated outbreak of lung injuries? Are there any airlines that allow electronic cigarettes? Are electronic cigarettes a healthier alternative to regular cigarettes?* Posts focused on vaping etiquette and vaping regulation. Stack Exchange questions were also increasing over time, from just four in 2009 to 192 in 2021. It seems that Stack Exchange is primarily for vapers, with some information on vaping as a smoking cessation tool, but limited anti-vaping content. Stack Exchange may not be ideal for dissuading youth from vaping. Examples of Wikipedia pages are *Electronic cigarette, Cloud-chasing, Construction of electronic cigarettes, Pax Labs, Regulation of electronic cigarettes*. Overall, Wikipedia pages centered on vape products and major events such as EVALI, detailing that vaping is framed as a consumer product requiring regulation, rather than an alternative to combustible cigarettes - perhaps indicative of Wikipedia's bias against vaping. Examples of wikiHow pages were *How to Fix Vape Pen Wires, How to Charge a Vape Pen, How to*

*Choose an Electronic Cigarette*. Broadly, wikiHow seems to cater to vape enthusiasts, with only a single anti-vaping article (*How to Stop Vaping*). wikiHow may thus not present balanced views on vaping and there were no pages on vaping as a tool for smoking cessation.

| | Reddit | Quora | Medium | Stack Exchange | Wikipedia |
|---|---|---|---|---|---|
| what is vaping for? | nicotine, smoking, juice, mods, flavor | use in quitting smoking, quit smoking, nicotine, harm reduction, lungs | your wellbeing, to prevent my nicotine addiction from killing me, smoking cessation aids, saving former smokers from smoking, substitute to smoke | satisfy your craving for a cigarette, smoking cessation aid, personal use, quitting smoking | use-associated injury, smoking cessation, quitting smoking, nicotine delivery device, harm reduction, giving up smoking |
| What are the advantages of vaping compared to smoking cigarettes? | your lungs will probably be a lot happier, you save a ton of money, you don't smell like nasty cigs and have stink, you can do it without reeking | your health and breath will improve, you don't stink and don't have brown teeth, you can try various flavors, very safe, without burning tobacco, 95% safer, less harmful, cheaper | without the fire, tars, smells or ash, they are not at all harmful, simpler and is used worldwide, safe and healthier, less harmful, healthier | reduce your intake of toxins greatly, healthier, fewer long term health effects, much safer, | smoking cessation, less harmful and more socially acceptable, more efficient |
| why are teens vaping? | trying to fit in, to quit smoking, to save money, to look good, anxiety | tobacco harm reduction, to show off to their friends, to quit or stay quit from smoking, to help with stress and anxiety, to fit in with their peers | to fit in, marketing, breaks society's rules, believe it's safe, quit smoking | think they are invincible, quitting smoking | to help them quit, to help them give up smoking, desire to smoke and withdrawal |
| what is the biggest concern with vaping? | youth vaping epidemic, your voice, your lungs, your build and voltage, your coils, your current | your lung, your health, you do not know what you are inhaling, worse on the lungs, vitamin e acetate | underage smoking, wattage, price, nicotine addiction, severe lung disease, lung injury | oral health, not safe, fire hazard, harmful vapor, fire and explosion risk, | underage people obtaining and getting addicted, smoking-related diseases, smoking, severe pulmonary disease, lung illness, lung disease, |

Table 1: Question-answering results for various platforms.

**Question Answering** We now present answers to four questions and up to 10 most representative answers across each platform in Table 1. Some questions provided less than 10 representative answers due to data availability. For the question *What is vaping for?*, we note answers framing vaping as a smoking cessation tool in Quora, Medium, and Stack Exchange, *harm reduction, to prevent my nicotine addiction from killing me, smoking cessation aid*. Reddit tended to frame vaping as a hobby, with answers such as *juice, mods, flavor*. Wikipedia had a mix of answers, some centered on EVALI, *use-associated injury*, and others on vaping as harm reduction *giving up smoking, smoking cessation, quitting smoking*. Regarding *What are the advantages of vaping compared to smoking cigarettes?*, all platforms provide answers around improved health and reduced cost, *your health and breath will improve, cheaper, your lungs will probably be a lot happier, reduce your intake of toxins greatly, less harmful and more socially acceptable*. For *Why are teens vaping?*, all platforms suggest that peer influence is a factor, *to show off to their friends, to fit in with their peers*. We note answers which may indicate that some teens vape as a form of harm reduction, *to quit smoking, to help them give up smoking*, in line with recent work [17]. Finally, for *What is the biggest concern with vaping?*, all platforms except Reddit indicate EVALI-related responses such as *vitamin e acetate, severe lung disease, severe pulmonary disease* or substance use concerns *underage people obtaining and getting addicted, nicotine addiction*. Reddit answers containe such concerns but also focused on the hobbyist aspects of vape devices, *your build and voltage, your coils, your current*. We conjecture that Quora, Medium, and Stack Exchange may be appropriate avenues for those seeking information around vaping as an alternative to smoking. Reddit may not be useful for those wanting to make the transition from smoking, and may even provide incorrect information to youth. Wikipedia contains a range of views, and may overemphasize vaping harms. We also note some EVALI-centric answers, which may exaggerate vaping risks.

**Cloze Tests** We use cloze tests to gauge the aggregate framing around vaping, across Reddit and Quora. Table 2 shows the cloze test results for several probes using Quora and Reddit data. Broadly, Reddit frames vaping as a hobby, with major concerns such as *cost, smell, money, temperature, flavor*. Quora has a greater focus on the health effects of vaping, with concerns *safety, taste, health, cost, flavor*. We note similar results when detailing *The worst thing about vaping is [MASK]*, where Reddit indicates *money, smell, smoking, convenience, efficiency*, while Quora provides *taste, stupidity, smell, safety, that*. Much like our other findings, Reddit may be ideal for vaping hobbyists, and Quora may be a more appropriate venue for those seeking to transition from smoking to vaping.

**Translation across Platforms through Large-scale Language Models** We demonstrate single word translation results from our large-scale language models to understand differences between Reddit and Quora. Upon manual inspection, we present misaligned pairs for Reddit and Quora posts, and illustrative sentence examples in Table 3. We first indicate the *fda, propaganda* pair which demonstrates how individuals on Reddit frame the FDA, which regulates vaping in the US, similar to how Quora users discuss anti-vaping propaganda. Such evidence may indicate that Reddit users frame the FDA as being anti-vaping, a stance often taken by vaping hobbyists. We then indicate the *ingredients, additives* pair. Reddit tends to use *ingredients*, which has a more neutral connotation, unlike Quora, which uses *additives*, with a more sinister connotation. In the Quora sentence examples, we note that *additives* is often mentioned with *dangerous, untested, contaminants*, unlike the equivalent *ingredients* in Reddit, which is mentioned alongside *safe, less harm*. It seems that Reddit takes a largely provaping stance, possibly not always evidence-based, and perhaps other platforms may provide a more balanced framing for those looking for information on vaping.

|  | Reddit (probability) | Quora (probability) |
|---|---|---|
| The main issue with vaping is [MASK] | cost (0.140) | safety (0.183) |
|  | smell (0.053) | taste (0.144) |
|  | money (0.044) | health (0.077) |
|  | temperature (0.040) | cost (0.074) |
|  | flavor (0.039) | flavor (0.032) |
| The worst thing about vaping is [MASK] | money (0.076) | taste (0.200) |
|  | smell (0.053) | stupidity (0.049) |
|  | smoking (0.041) | smell (0.036) |
|  | convenience (0.040) | safety (0.036) |
|  | efficiency (0.035) | that (0.032) |
| Teens like vaping because it's [MASK] | cool (0.173) | cool (0.461) |
|  | safe (0.063) | safe (0.048) |
|  | convenient (0.049) | popular (0.043) |
|  | new (0.035) | dangerous (0.032) |
|  | fun (0.034) | enjoyable (0.025) |
| Vaping is [MASK] to our health. | bad (0.479) | dangerous (0.312) |
|  | harmful (0.200) | detrimental (0.229) |
|  | dangerous (0.068) | harmful (0.229) |
|  | detrimental (0.050) | bad (0.68) |
|  | important (0.035) | damaging (0.064) |

Table 2: The top five candidate words ranked by BERT probability for the cloze test "*The main issue with vaping is [MASK]*", "*The worst thing about vaping is [MASK]*", "*Teens like vaping because it's [MASK]*", "*Vaping is [MASK] to our health.*" for Reddit and Quora data.

## 4  Discussion

**Implications of findings** Our RQ was to explore how vaping is framed across online knowledge dissemination platforms. A strength of our work is how the different techniques we applied validate each other as well as reveal differences across platforms. For example, cloze tests and question answering results indicate that Quora is an appropriate venue for those looking to transition from smoking to vaping. Reddit and wikiHow may be for vaping hobbyists and may not sufficiently dissuade youth vaping. Conversely, Wikipedia may exaggerate vaping harms, dissuading smokers from transitioning.

**Recommendations** Key to how vaping is framed is the inclusion of vaper viewpoints when writing articles on vaping. Where possible, vapers themselves should be consulted on articles about vaping. For example, a panel staffed by vapers can comment on vaping-related questions and answers on Quora, providing suggestions on how answers can more accurately represent vaper concerns. As the vaping landscape continues to evolve, it is possible that more vaping regulation inimical to smoking cessation is proposed. We suggest vaping regulation that clarifies the role of vapes as a smoking cessation tool. To improve framing around vaping, minimize marginalization, and possibly mitigate vaping (mis)perceptions online, stakeholders can design informational tools which reinforce or mitigate vaping (mis)perceptions. An example tool can use brief exposure to evidencebased information about vapes, perhaps reducing vaping misperceptions. Such interventions may shift the beliefs of those against vaping, thereby reducing stigma around vaping for those who smoke and want to make a transition. **Limitations** Our findings relied on the validity of data collected with our search terms. We used a range of established techniques to search for all articles/posts relevant to vaping, and our data contained text aligned with how vaping is framed. We are thus confident in the comprehensiveness of our data. We note that the recall of the search string was not tested. We note that our data may not be generalizable to how vaping is framed globally. We were not able to obtain statistics about how many times an article was read or shared. Findings may also not apply to other related issues that are also heavily politicized (e.g., abortion) or other contexts (e.g., vaping frames in Europe). We also note the limitations of BERT, such as its inability to learn in few-shot settings.

| Misaligned Pairs | Reddit | Quora |
|---|---|---|
| **Illustrative examples** | | |
| <fda, propaganda> | thats why fda says there is no risk with getting addicted to nicotine replacement therapy | Even with all the propaganda about vaping, there is scientific studies that suggest nicotine IS NOT nearly as addictive as it is portrayed to be |
| | The fda would rather you smoke than vape anyway | Seriously, nearly 20 years of repeated science on the topic of OIL FUMES trashing lungs! Not just vaping and doesn't have a damn thing to do with nicotine! Don't fall for the Reefer Madness propaganda of the anti-vaping stupidity and ignorance! |
| <ingredients, additives> | Vaping definitely does way less harm than cigarettes, just read the ingredients and the level of formaldehyde has already been proven to be perfectly safe to breathe in | Vaping is shaping up to be significantly more dangerous than cigarette smoking for a number of reasons, largely related to the quality of the stuff you are inhaling, the untested additives, and the concentration of chemicals |
| | The ingredients that you put inside are Propylene Glycol (PG), Vegetable Glycerin (VG), nicotine by mg increments, and flavorings | It usually contains propylene glycol, glycerin, nicotine, flavorings, additives, and contaminants |

Table 3: Misaligned word pairs and illustrative sentence examples for Reddit and Quora regarding vaping framing.